\documentclass[conference]{IEEEtran}

\usepackage{citesort}
\usepackage{amsmath}
\usepackage{amssymb}
\usepackage{amsthm}
\usepackage{graphicx}
\usepackage{url}
\usepackage{textcomp}
\usepackage{xcolor}
\usepackage{multirow}
\usepackage{flushend}
\usepackage{booktabs}
\usepackage{siunitx}
\def\BibTeX{{\rm B\kern-.05em{\sc i\kern-.025em b}\kern-.08em
    T\kern-.1667em\lower.7ex\hbox{E}\kern-.125emX}}

\newlength\figurerowsep
\makeatletter
 \@ifundefined{Hline}{}
\makeatother

\title{Packet Interval Modulation-based NFC for Scalable IoT around Meandered E-textile}

\author{
  \IEEEauthorblockN{
    Shigemi Ishida\IEEEauthorrefmark{1}\IEEEauthorrefmark{3},
    Ryo Takahashi\IEEEauthorrefmark{2}\IEEEauthorrefmark{3},
    and Satoshi Ido\IEEEauthorrefmark{1}\IEEEauthorrefmark{3}
  }
  \IEEEauthorblockA{
    \IEEEauthorrefmark{1}School/Grad School of Systems Information Science, Future University Hakodate, Japan\\
    Email: \{ish, g2125004\}@fun.ac.jp
  }
  \IEEEauthorblockA{
    \IEEEauthorrefmark{2}School of Engineering, The University of Tokyo, Japan\\
    Email: takahashi@akg.t.u-tokyo.jp
  }
  \thanks{\IEEEauthorrefmark{3}These authors contributed equally to this work.}
}

\IEEEoverridecommandlockouts\IEEEpubid{\makebox[\columnwidth]{ \copyright~2026 IEEE \hfill} \hspace{\columnsep}\makebox[\columnwidth]{}}

\begin{document}
\maketitle

\begin{abstract}
Near-field communication (NFC) sensor tags are promising for low-cost, scalable environmental monitoring due to their battery-free, maintenance-free operation.
Recent advances in NFC reader garments, integrated with conductive textile coils, allow users to interact seamlessly with tags scattered across their daily environment. 
However, dynamically identifying and reading out a fluctuating number of tags remains challenging.
If multiple tags speak and collide, they must randomly retry until an empty slot is found. 
Standard NFC identifies tags using time-slotted ALOHA, which assigns tags to fixed time slots.
If multiple tags speak and collide, they must retry with a tag filter to avoid collision until all tags are identified.
This rigid setup is fragile in dynamic settings; as tags enter and leave the communication range, the heavy overhead of slot re-negotiation and inevitable collisions restricts data throughput.
Here, we propose an asynchronous NFC based on packet interval modulation. 
By treating minimal ID-only packets as pulses and encoding data in their intervals, we enable communication through silence.
This ensures the channel remains mostly unoccupied, drastically reducing collisions even in high-density environments.
Our simulations show that for ten tags, our method achieves a 4-fold increase in throughput compared to the ALOHA approach. 
Furthermore, the system remains robust, maintaining over 160 bps/tag throughput even under extreme fluctuations in the number of tags over 20.
\end{abstract}


\IEEEpeerreviewmaketitle

\section{Introduction}
\label{sec:intro}

\begin{figure}[t!]
  \centering
  \includegraphics[width=1\linewidth]{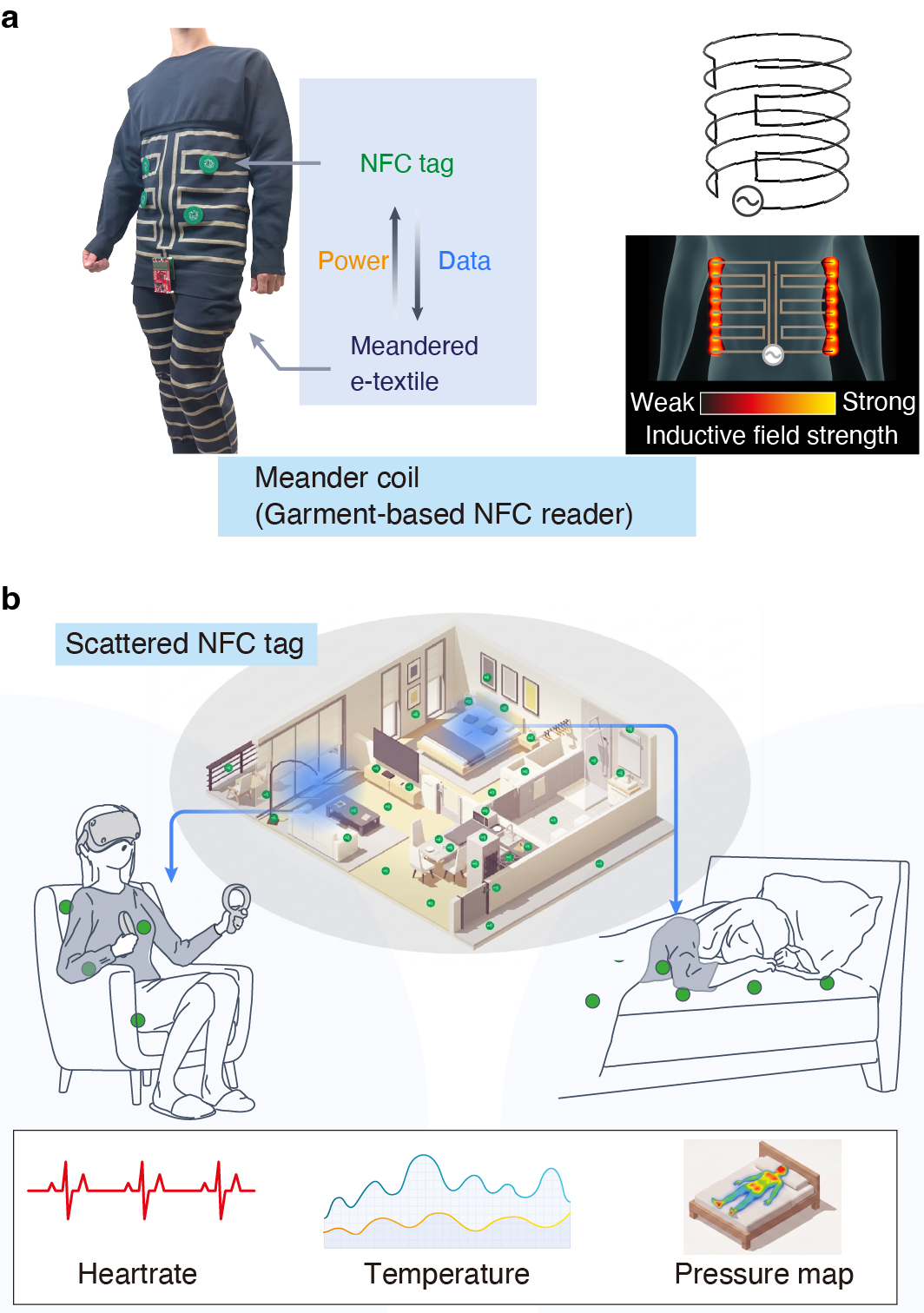}
  \caption{Concept of the garment-based NFC sensing platform in smart environments. (a) Overview of meandered e-textiles for scattered NFC tags. (b) The wearer acts as a mobile probe, seamlessly acquiring health and environmental data by interacting with multiple NFC tags scattered across household items and furniture during daily activities.}
  \label{fig:overview}
\end{figure}

Near-field communication (NFC)-based sensor networks can provide a battery-free, maintenance-free platform for on-demand environmental sensing.
Recent advances in NFC reader garments have established physical hardware for this vision by integrating stretchable meander coils directly into clothing~\cite{zhu_robust_2024,takahashi_meander_2022}. 
Unlike traditional loop or helical coils, these meander patterns provide a unique magnetic field confinement effect that concentrates magnetic flux precisely at the textile surface.
This confinement maintains even as the coil scales to large, body-sized dimensions, maximizing inductive coupling with nearby NFC tags while minimizing external electromagnetic or body interference.
Such a garment-style NFC reader enables users to act as mobile probes, seamlessly interacting with multiple NFC tags scattered across furniture or household items simply by being in proximity in daily life (see Fig.~\ref{fig:overview}). 
However, dynamically identifying and reading a fluctuating number of tags in real time remains challenging for existing protocols.
Standard NFC identifies tags using time-slotted ALOHA, which uses rigidly synchronized transmission windows~\cite{noauthor_cards_nodate}. 
This centralized coordination is fragile in dynamic settings; as tags enter and leave the range during user movement, the heavy overhead of slot re-negotiation and inevitable data collisions severely restrict throughput.

Existing strategies for high-density NFC sensing in the HF band typically operate under the assumption of static tags and readers~\cite{han_battery-free_2018,han_lin_yeap_compact_2025,zou_nfcrfid-enabled_2025}. 
A common limitation of these systems is the requirement for pre-registering the unique IDs of all NFC tags in use~\cite{han_battery-free_2018}.
For example, in scenarios where the reader and tags remain stationary, such as hospital beds composed of one NFC reader-integrated bed and 65 NFC tags mounted on the patient's skin, data is obtained sequentially according to a pre-defined inventory. 
In more complex scenarios where a mobile NFC reader needs to scan a varying number of tags, spatial optimization techniques—such as K-means clustering—have been employed to group tag locations~\cite{han_lin_yeap_compact_2025}.
However, these approaches cannot deal with in-the-wild scenarios where the system encounters unregistered tags or unpredictable changes in the tag population.
In UHF-band RFID networks, scalability is even more challenging, leading to the development of decentralized or inventory-free MAC protocols that can handle dynamic tag information without prior knowledge of their IDs~\cite{sangar_wichronos_2020,xie_enabling_2024}.
One example is packet interval modulation in which information is encoded in the timing between packets rather than the packet content itself~\cite{zhu_challenges_2005,sangar_wichronos_2020}.
By using extremely short pulses as packets, this asynchronous protocol minimizes channel occupancy to near-zero, effectively avoiding collisions without the need for centralized coordination.

Building on packet interval modulation, this paper proposes a scalable, asynchronous NFC scheme. 
Unlike standard slotted methods, our approach is asynchronous, utilizing the precise timing of pulses to represent data without the need for rigid synchronization or centralized slot management. 
By removing the overhead of slot re-negotiation, the reader garment can instantly recognize and acquire data from tags even as their density changes. 
Unlike previous NFC anti-collision approaches, the asynchronous packet interval enables scalable, energy-efficient data transfer against tens of NFC tags (see Tab.~\ref{tab:protocol_comparison}).
Our simulations demonstrate that for ten tags, our method achieves a $4$-fold increase in throughput compared to the standard dynamic ALOHA approach. 
Furthermore, the system proves robust in extreme conditions, maintaining over 160 bps/tag throughput during rapid fluctuations in tag counts.
These results provide a scalable foundation for high-density, invisible wearable healthcare integrated into smart environments.

\begin{table}[t!]
\centering
\caption{Comparison of NFC multi-tag identification protocols.}
\label{tab:protocol_comparison}
\begin{tabular}{@{}llll@{}}\toprule
 \textbf{Method} & \textbf{Sync} & \textbf{Tag's overhead} & \textbf{Tag number} \\ \midrule
 Slotted ALOHA~\cite{noauthor_cards_nodate}  & Synchronous & High (Slotting) & $<5$ \\
K-means cluster~\cite{han_lin_yeap_compact_2025} & Synchronous & Medium & $<10$ \\
\textbf{Packet interval} & \textbf{Asynchronous} & \textbf{Minimal} & $<20$ \\  \bottomrule
\end{tabular}
\end{table}

\begin{figure}[t!]
  \centering
  \includegraphics[width=1.0\linewidth]{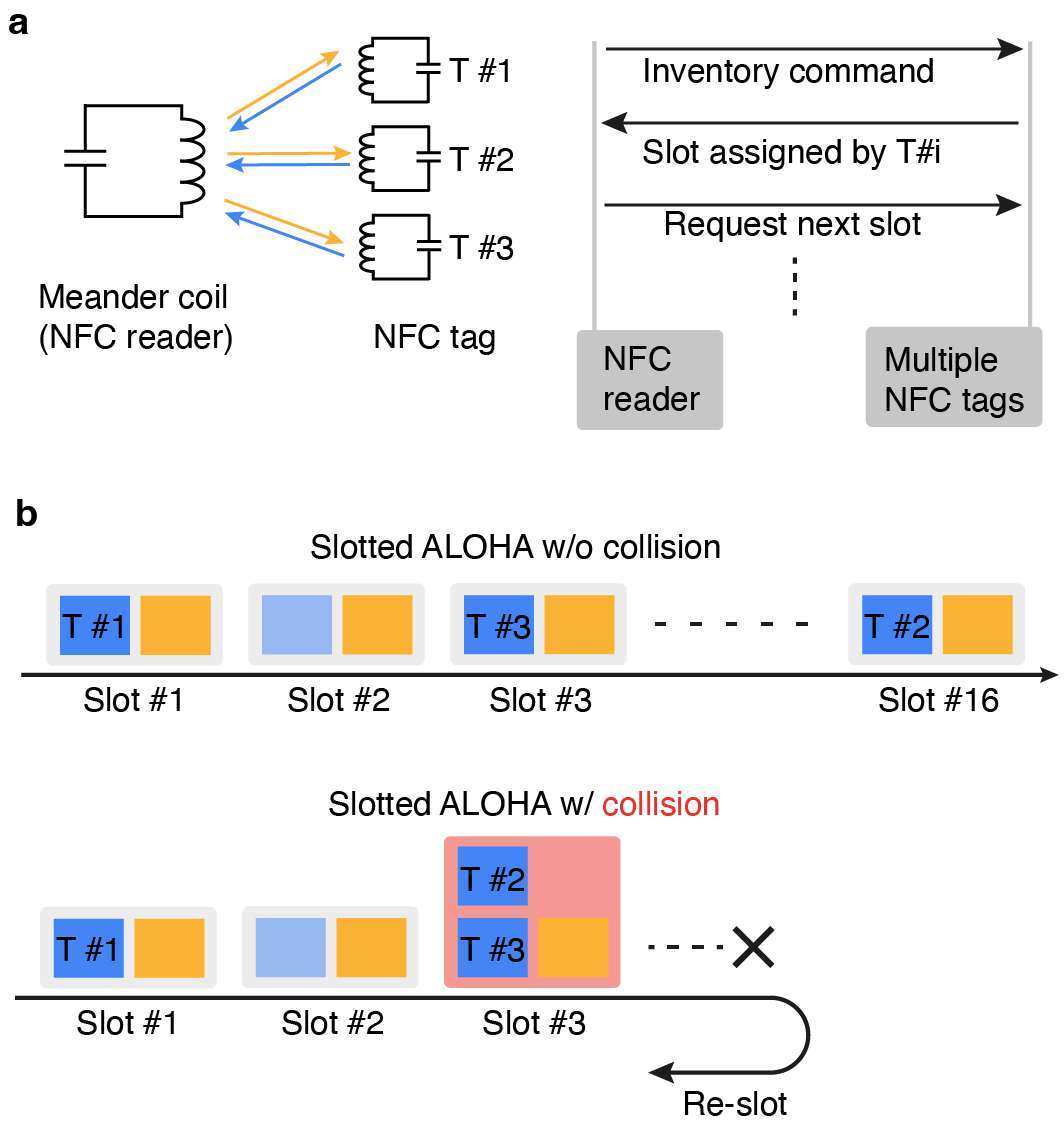}
  \caption{Slotted ALOHA mechanism in standard NFC. (a) Signal flow: The reader synchronizes tags via inventory commands, where each tag responds in a randomly assigned slot. (b) Collision occurrence: When multiple tags ($T\#2$ and $T\#3$) select the same slot, a collision occurs, forcing a time-consuming re-slot process. This rigid synchronization limits performance in high-density or dynamic tag scenarios.}
  \label{fig:aloha}
\end{figure}

\section{Packet interval modulation for NFC}
\label{sec:design}

\subsection{System overview}

Standard NFC communication is designed for one-on-one interactions over very short distances of \qtyrange[range-phrase = --,range-units  = single]{1}{2}{\cm}, such as a smartphone tapping a payment terminal or a single card unlocking a door.
To ensure the reader talks to only one tag at a time, NFC relies on a protocol called Slotted ALOHA, where a central reader organizes time into discrete slots and tells tags when to speak (see Fig.~\ref{fig:aloha}a).
Each tag picks a slot based on its ID to send its data.
While this works perfectly for a single device such as a smartphone interacting with a single terminal, it becomes a major bottleneck for our vision of a wearable sensing platform.
Our garment-style reader focuses on interacting with dozens of scattered tags simultaneously across a wide garment area.
In such a high-density IoT scenario that the original protocol never anticipated, if multiple tags pick the same slot, they collide and the data is lost (see Fig.~\ref{fig:aloha}b). The reader must then stop and restart a long negotiation process to reorganize the tags. 
Basically, when 5 to 10 or more tags are present, the probability that multiple tags pick the same slot increases dramatically, leading to constant data collisions.
These constant collisions create a slow, unreliable connection that fails to capture the data from the many tags the user encounters in a high-density IoT scenario.

\begin{figure}[t!]
  \centering
  \includegraphics[width=1\linewidth]{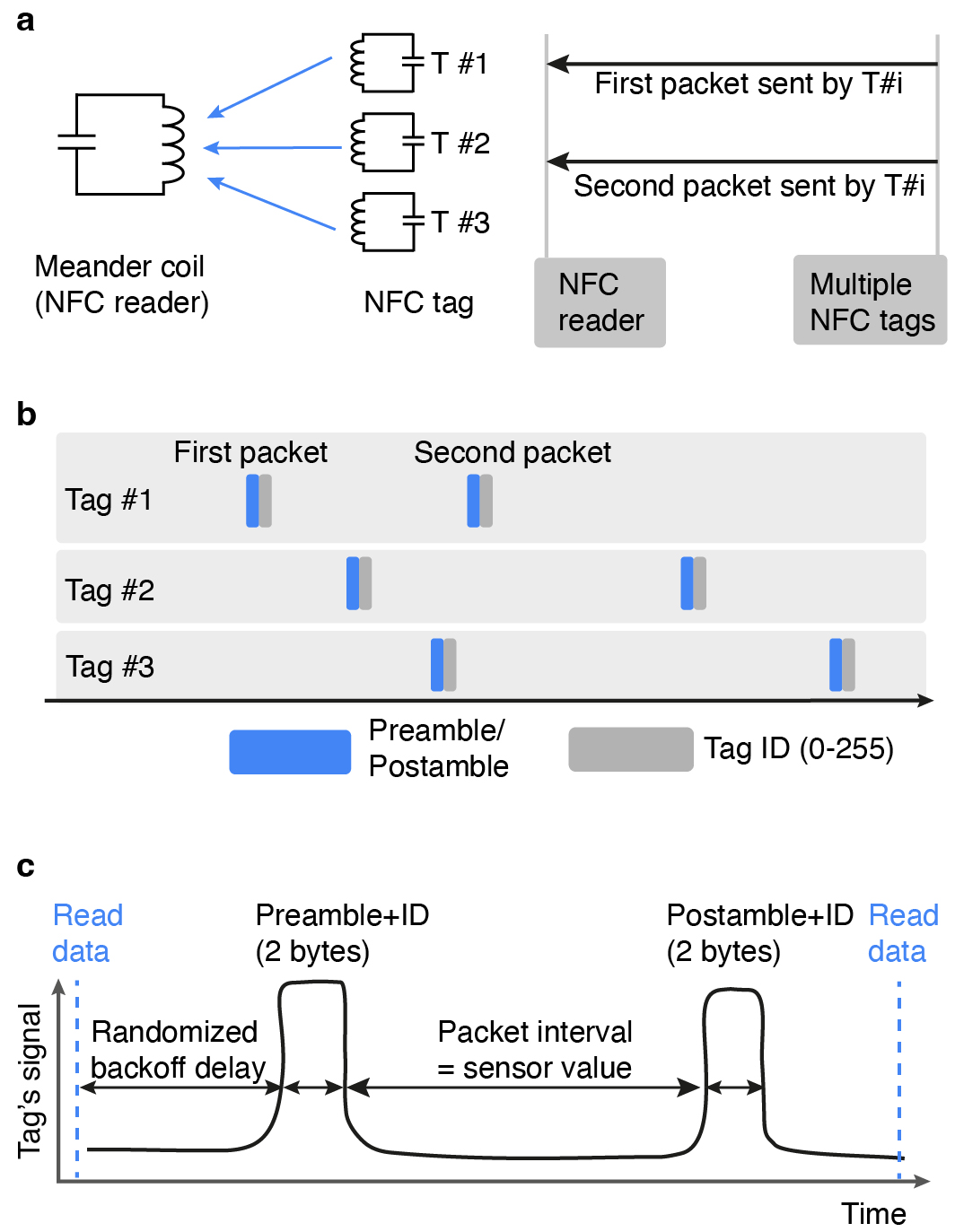}
  \caption{Overview of packet interval modulation in our proposed NFC architecture. (a) Unidirectional communication: Tags transmit data autonomously without waiting for reader commands. (b) Pulse-based multiplexing: High-speed, short pulses from multiple tags are distributed asynchronously, drastically reducing collision probability. (c) Data encoding: Sensor values are represented by the time interval between a preamble pulse and a postamble pulse, each containing a 1-byte Tag ID for identification.}
  \label{fig:pim}
\end{figure}

To overcome the collisions and synchronization overhead of standard NFC, we propose an asynchronous NFC scheme based on packet interval modulation (see Fig.~\ref{fig:pim}).
Unlike traditional protocols where the reader manages every transmission, our approach allows tags to transmit data autonomously, as shown in Fig.~\ref{fig:pim}a.
The reader simply provides a continuous magnetic field, and tags broadcast short, independent pulses whenever they update their sensor values.
The fundamental principle of PIM is to encode information not within the bits of a long packet, but in the temporal distance between two short pulses (see Fig.~\ref{fig:pim}b).
Each packet sent by a tag is extremely brief, consisting of only a few bytes for synchronization and identification.
Because these packets are temporally sparse, the probability of multiple tags transmitting at the exact same millisecond is statistically near zero.
This allows tags $T\#1$, $T\#2$, and $T\#3$ to coexist on the same frequency without a central coordinator.
As detailed in Fig.~\ref{fig:pim}c, each data transmission cycle consists of a pair of pulses: the first pulse including preamble and ID, and the second pulse including postamble and ID.
By reducing the on-air time of each tag to these minimal pulses, the system achieves a massive reduction in channel occupancy. 
This decentralized architecture allows the wearable reader to capture sensor data from a varying number of tags instantly and continuously.

\subsection{Symbol design}

We design our packet interval modulation by optimizing the physical layer of the NFC Type-V standard (ISO/IEC 15693) with the frequency range of \qty[separate-uncertainty = true, multi-part-units = single]{13.56 \pm 0.424}{\MHz}. 
Because standard Type-V is mainly designed in the case of weak magnetic coupling between coils, NFC for e-textiles is also compatible because the NFC reader embedded into the textiles needs to read out the patch-sized small NFC tags on the skin via weak inductive coupling~\cite{zou_nfcrfid-enabled_2025}.
The integration of packet interval modulation into the Type-V needs to minimize the on-air time of each transmission through three main optimizations of data rate, packet size, and pulse modulation.

The first step is to increase the data rate. 
While the standard Type-V is limited to \qty{26.48}{kbps} to prioritize signal robustness over long distances, our system shifts to a high-speed \qty{424}{kbps} subcarrier frequency.
This transition is possible because our asynchronous architecture eliminates the need for reader-to-tag downlink commands, which traditionally consume significant bandwidth for synchronization and slot management. 
By removing this overhead, we can allocate the maximum available bandwidth to the tag-to-reader uplink.
Previous research on meandered e-textiles confirms this link is reliable: even at \qty{424}{kbps} BPSK with a \qty{0}{dBm} reader output, the system maintains a BER of \num{1e-5} within a \qty{1}{cm} distance~\cite{takahashi_full-body_2025}.

\begin{figure*}[t!]
  \centering
  \includegraphics[width=2\columnwidth]{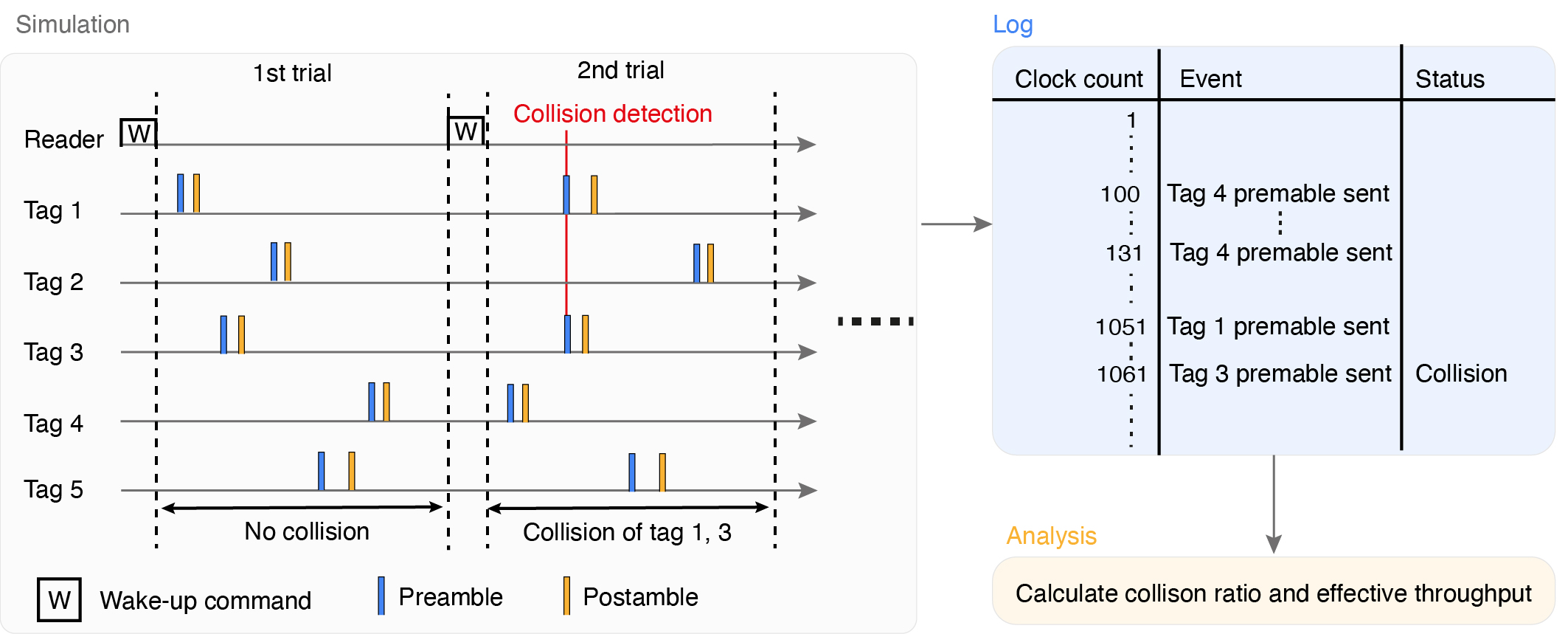}
  \caption{Example of our simulation timeline of the packet interval modulation when the tag number is five.}
  \label{fig:simulation}
\end{figure*}

The second step is to minimize the packet length. 
To make the transmission time shorter, we condense each data packet into a tiny 2-byte frame consisting of only a synchronization pulse—either a preamble or postamble—and a 1-byte unique Tag ID, similar to WiChronos~\cite{sangar_wichronos_2020}. 
This simplified structure allows the system to identify up to 256 ($=2^8$ bit) unique tags while maintaining minimal on-air time.

The third step optimizes pulse modulation timings, focusing on randomized backoff delay and value-to-interval mapping. 
The former is a random interval between the completion of sensor data readout and the initiation of the tag’s first preamble, while the latter encodes digital sensor values into the temporal spacing between pulses. This approach presents a trade-off: shorter delay intervals minimize latency but exacerbate packet overlaps in high-density scenarios. 
Temporal instabilities, such as clock drift in the NFC components also constrain accuracy. 
To balance these factors, we employ numerical simulations to identify the optimal timing parameters.
However, in our NFC-based system, such drift is intrinsically minimized because the tag synchronizes its internal clock to the reader’s stable carrier frequency (\qty{13.56}{\MHz}) via PLL or equivalent clock recovery. 
While this synchronization provides sub-microsecond precision, minor timing jitters may still arise from tag-side circuit variations.

\section{Numerical Simulation}

\subsection{Experimental setup}

To evaluate the performance of our proposed asynchronous packet interval modulation scheme, we conducted numerical simulations using a custom Python-based framework \textcolor{blue}{(\url{https://github.com/yokota-cronos/2026_globecom_pim_for_NFC})}, as shown in Fig.~\ref{fig:simulation}. 
The simulation models the interaction between a large-scale wearable reader and a variable number of $N$ NFC tags ($N = 1$ to $20$) at $100000$ trials.
Each tag is programmed to transmit $8$-bit sensor data at a fixed update rate, with its start pulse randomized within a window to simulate asynchronous behavior. 
We first validated our simulation model by comparing the numerical results with theoretical collision probabilities. 
Then, we investigated the optimization of pulse modulation timing to minimize collisions and a stable data cycle. 
We specifically analyzed the impact of the randomized backoff window by varying its size.
Lastly, we examined the value-to-interval mapping strategy to determine how the encoding of sensor data affects communication robustness.
As for the simulation environment, we developed a discrete-event simulator operating at the subcarrier clock level. 
While general tools like ns-3 provide complex physical layer modeling considering random bit errors or jitter, they are unnecessary for our specific hardware.
The main reason is that our system uses a \qty{13.56}{\MHz} carrier with meander-coil coupling, which ensures an extremely low BER owing to the strong inductive confinement through the meander coil and stable clock synchronization due to the low carrier frequency.
This allows the simulation to focus entirely on pulse collisions and timing overlaps, which are the main issues with our system scalability.
In total, we set up that the reader triggers a wake-up signal every \qty{100}{\ms}, within which each tag completes its randomized backoff and data transmission within \qtyrange{30}{50}{ms} to avoid inter-cycle collisions.

\subsection{ALOHA vs Packet Interval Modulation}

\begin{figure*}[t!]
  \centering
  \includegraphics[width=2\columnwidth]{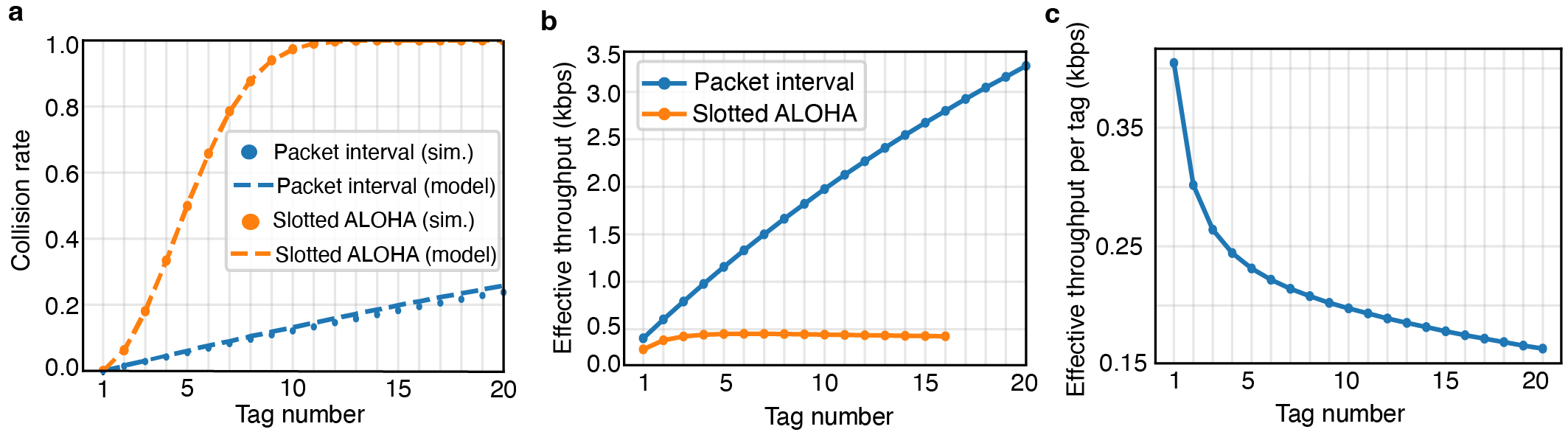}
  \caption{Simulation results of the proposed packet interval modulation compared with the conventional slotted ALOHA: (a) Collision rate and (b) total effective throughput of the proposed packet interval and slotted ALOHA. (c) Effective throughput per tag for the proposed packet interval.}
  \label{fig:comparison}
\end{figure*}

First, we compare the reliability of our packet interval modulation with the standard slotted ALOHA system.
In a slotted ALOHA system, we consider the probability that at least one collision occurs among $N$ tags competing for $L$ available slots. 
We define the system-level collision probability $P_{\rm ALOHA}$ as the probability that any of the slots chosen by the $N$ tags contain more than one transmission. 
Similar to the Birthday Paradox, $P_{\rm ALOHA}$ is mathematically similar to the complement of all tags selecting different slots, expressed as:
\begin{align}
    P_{\rm ALOHA} &= 1 - P_{\rm success\_ALOHA} \\
                  &= 1 - \cfrac{^L P_N}{L^N} \\
                  &= 1 - \cfrac{L!}{L^N (L-N)!}
\end{align}
where $P_{\rm success\_ALOHA}$ represents the probability that every tag randomly selects a unique slot and $^L P_N$ denotes the number of permutations for choosing.
As $N$ approaches $L$, this probability rapidly converges to $1$, demonstrating the rigid scalability limits of slot-based coordination.
The Type-V protocol typically utilizes an inventory command that defines either 1 or 16 slots for anti-collision resolution, which means the Type-V supports up to 16 NFC tags.
When the collision happens, the reader initiates another inventory round using a mask to quiet the tags that have already been successfully identified, to efficiently assign the collided tags in the remaining slots.
However, as we will describe later, when the number of tags $N$ exceeds 10, the probability of collisions within only 16 slots increases dramatically.

In contrast, the packet interval modulation operates in continuous time, where a collision occurs only if any of the pulses within a transmission sequence overlap with those of another tag. 
We first calculate a collision probability when there are two tags.
Preamble and postamble are pulses whose duration is $\tau$ of a single 16-bit packet ($\approx \qty{75.5}{\mu s}$ at \qty{424}{kbps} Manchester coding).
When two tags transmit a single pulse independently within a randomized backoff window $T_{\rm backoff}$, a pulse transmission from one tag is successful only if the other tag transmits a pulse outside of the vulnerable period of $2\tau$ (the duration of the pulse itself plus the time during which a preceding pulse would still overlap).
The probability of a single pulse colliding with a pulse from the other tag is given by:
\begin{align}
 \hat{P}_{\rm coll} = \frac{2\tau}{T_{\rm backoff}}
\end{align}

When two tags transmit two pulses of preamble and postamble, four types of collisions may occur:
C1)~preamble--preamble,
C2)~preamble--postamble,
C3)~postamble--preamble,
and C4)~postamble--postamble.
When each tag sends random data, the duration between preamble and postamble is random within a window depending on the data bits.
The probability of each collision type is then given by $\hat{P}_{\rm coll}$.
If the collision probability is sufficiently small, i.e., $\hat{P}_{\rm coll} \ll 1$, we can assume collision scenarios C1--C4 are independent.
This assumption is natural because two types of collisions simultaneously occur with the probability in the order of $\mathcal{O}({\hat{P}_{\rm coll}}^2)$.
The probability of successful two-pulse transmission in a two-tag scenario is therefore given by:
\begin{align}
 P_{{\rm success\_PIM},2} &= (1 - \hat{P}_{\rm coll})^4 \\
                     & \approx 1 - 4\hat{P}_{\rm coll}
                       \quad (\hat{P}_{\rm coll} \ll 1)
\end{align}

We extend this collision model to the scenario with a larger number of tags.
Let $N$ be the number of tags.
Two-pulse transmission from a specific tag is successful only when both two pulses avoid collision with pulses from other $(N-1)$ tags.
Thus, the system-level collision probability $P_{\rm PIM}$ is derived:
\begin{align}
 P_{\rm PIM} &= 1 - {P_{{\rm success\_PIM},2}}^{N-1} \\
             &= 1 - \left(1 - \frac{8\tau}{T_{\rm backoff}}\right)^{N-1}
\end{align}

We conducted numerical simulations to validate these theoretical models. 
We varied the number of tags $N$ from \num{1} to \num{20} and compared the collision rates and calculated the effective throughput of the packet interval modulation and slotted ALOHA.
In our simulation, we focused on environmental sensing applications where data is typically updated every \qty{100}{ms}. 
To balance low latency with high reliability, we set the randomized backoff window $T_{\rm backoff}$ to approximately \qty{38.6}{\ms} (corresponding to $2^{14}$ clock units). 
This allows for a sufficiently sparse temporal distribution while maintaining a fast update cycle suitable for real-time human-centric monitoring.
The results show that while the slotted ALOHA system (with $L=16$ slots) suffers from a rapid increase in collisions as $N$ exceeds \num{10}, the packet interval scheme maintains a significantly higher success rate (see Fig.~\ref{fig:comparison}a). 
In terms of effective throughput, packet interval outperforms ALOHA by approximately $4$-fold increase in high-density environments, in addition to maintaining stable effective throughput per tag over approximately \qty{0.1}{kbps} even when the tag number reaches \qty{20} (see Fig.~\ref{fig:comparison}b-c).
These low collision rate and stable effective throughput per tag are sufficient for human-centric on-demand environmental monitoring with scattered NFC sensor tags.

\subsection{Optimal pulse modulation for tag number}

\begin{figure}[t!]
  \centering
  \includegraphics[width=1\columnwidth]{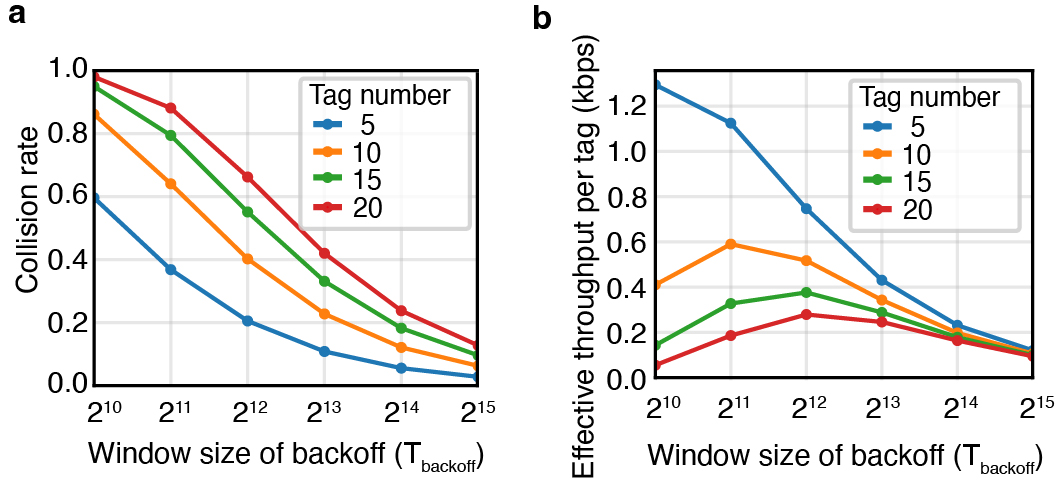}
  \caption{Performance dependence on the randomized backoff window size ($T_{\rm backoff}$): (a) Collision rate and (b) effective throughput per tag for different $T_{\rm backoff}$).}
  \label{fig:backoff}
\end{figure}

We designed the temporal parameters of our protocol to maximize network capacity while maintaining a low collision rate.
The most important factor in this optimization is the duration of the randomized backoff window, $T_{\rm backoff}$, which decides the overall traffic density. While a longer backoff window reduces the likelihood of pulse overlaps, it simultaneously lowers the effective throughput and increases latency. 
To determine the optimal balance, we conducted simulations varying $T_{\rm backoff}$ from $2^{10}$ to $2^{15}$ clock units (i.e., \qtyrange{2.4}{77.2}{ms}), as shown in Fig.~\ref{fig:backoff}.
To support real-time environmental monitoring, we defined a target update cycle of \qty{100}{\ms}. 
Since the tags lack a feedback mechanism to confirm successful reception by the reader, we adopted a strategy of transmitting the same packet twice to enhance communication reliability.
Although multiple retransmissions increase power consumption, our simulation revealed that sending the data exactly twice with a backoff window of $2^{14}$ clock units (\qty{38.6}{\ms}) achieves a collision rate of less than \qty{5}{\%} even with $20$ tags. Under this configuration, the system maintains an effective throughput over \qty{160}{bps}, ensuring that all 8-bit sensor data is reliably updated and transmitted within each \qty{100}{\ms} window.
Therefore, we use  $2^{14}$ as the optimal backoff window size.

\begin{table}[t!]
\centering
\caption{Impact of sensor value mapping strategies on system performance ($N=20$). The values in parentheses denote the collision rate and effective throughput per tag.}
\begin{tabular}{lccc}
\toprule
           & x1 wide interval             & x2 wide interval            & x3 wide interval            \\ \midrule
Gaussian & (0.23, 167 bps) & (0.24, 162 bps) & (0.25, 160 bps) \\ 
Random     & (0.24, 163 bps) & (0.25, 160 bps) & (0.25, 158 bps) \\
\bottomrule
\end{tabular}
\label{tab:random_redundancy}
\end{table}

We also investigated how the mapping strategy from sensor values to time intervals affects system robustness. 
The first technical concern is that clock frequency mismatches between the NFC reader and NFC tags can introduce errors when converting the measured pulse width back into sensor data.
The second concern is the risk of collisions when multiple tags observe identical environmental conditions (e.g., stable room temperature) and attempt to transmit identical intervals simultaneously.
To consider these two issues, we evaluate a redundancy factor (x1 to x3) to widen the pulse intervals, providing a buffer against such timing inaccuracies. 
Furthermore, to address the risk of persistent collisions when multiple tags observe near-identical environmental states, we compare a scenario where sensor values follow a concentrated Gaussian distribution ($\mu=128, \sigma=1$) against a uniform random distribution (0--255).
As summarized in Tab~\ref{tab:random_redundancy} ($N=20$, $T_{\rm backoff}=2^{\rm 14}$), our results indicate that both the collision rate and effective throughput remain almost constant at approximately \qty{0.24} and \qty{160}{bps}, respectively, because $T_{\rm backoff}$ is sufficiently longer than the pulse intervals.

Through the above simulation, we optimized the pulse interval modulation for NFC-based asynchronous sensing.
By selecting a backoff window of $2^{14}$ clock units (\qty{38.6}{\ms}) and incorporating the redundancy factor x2 for pulse intervals, we established a robust communication foundation. 
The dual-transmission strategy ensures high reliability with a collision rate around \qty{5}{\%} and an effective throughput per tag exceeding \qty{160}{bps} for up to 20 tags within a \qty{100}{\ms} update cycle.

\section{Conclusion}
\label{sec:concl}

This paper proposed a scalable and asynchronous NFC sensing scheme based on packet interval modulation to overcome the collision and overhead limitations of conventional slotted ALOHA in dynamic IoT environments. 
Through numerical simulations, we optimized the protocol parameters, showing a stable throughput exceeding \qty{160}{bps} for up to 20 tags within a \qty{100}{\ms} update cycle. 
These results confirm that the proposed method provides a dependable and high-capacity communication foundation for real-time environmental monitoring using scattered NFC sensor tags, outperforming traditional protocols by approximately four-fold in high-density scenarios.

\section*{Acknowledgment}

This work was supported in part by JST JPMJCS25N4 JPMJPR2515, and JSPS KAKEN 22K21343.

\bibliographystyle{IEEEtran}
\bibliography{references}

@misc{takahashi_full-body_2025,
	title = {Full-body {NFC}: body-scale near-field sensor networks with machine-knittable meandered e-textiles},
	shorttitle = {Full-body {NFC}},
	url = {http://arxiv.org/abs/2503.13240},
	doi = {10.48550/arXiv.2503.13240},
	urldate = {2026-04-28},
	publisher = {arXiv},
	author = {Takahashi, Ryo and Han, Changyo and Yukita, Wakako and Ho, John S. and Sasatani, Takuya and Noda, Akihito and Yokota, Tomoyuki and Someya, Takao and Kawahara, Yoshihiro},
	month = mar,
	year = {2025},
	note = {arXiv:2503.13240 [cs]},
}

@inproceedings{zhu_challenges_2005,
	address = {New York, NY, USA},
	series = {{MobiCom} '05},
	title = {Challenges: communication through silence in wireless sensor networks},
	isbn = {978-1-59593-020-0},
	shorttitle = {Challenges},
	url = {https://dl.acm.org/doi/10.1145/1080829.1080844},
	doi = {10.1145/1080829.1080844},
	urldate = {2026-04-13},
	booktitle = {Proceedings of the 11th annual international conference on {Mobile} computing and networking},
	publisher = {Association for Computing Machinery},
	author = {Zhu, Yujie and Sivakumar, Raghupathy},
	year = {2005},
	pages = {140--147},
}

@article{zhu_robust_2024,
	title = {A robust near-field body area network based on coaxially-shielded textile metamaterial},
	volume = {15},
	copyright = {2024 The Author(s)},
	issn = {2041-1723},
	url = {https://www.nature.com/articles/s41467-024-51061-x},
	doi = {10.1038/s41467-024-51061-x},
	language = {en},
	number = {1},
	urldate = {2026-03-31},
	journal = {Nature Communications},
	publisher = {Nature Publishing Group},
	author = {Zhu, Xia and Wu, Ke and Xie, Xiaohang and Anderson, Stephan W. and Zhang, Xin},
	month = aug,
	year = {2024},
	pages = {6589},
}

@article{zou_nfcrfid-enabled_2025,
	title = {{NFC}/{RFID}-enabled wearables and implants for biomedical applications},
	volume = {11},
	copyright = {2025 The Author(s)},
	issn = {2055-7434},
	url = {https://www.nature.com/articles/s41378-025-01010-5},
	doi = {10.1038/s41378-025-01010-5},
	language = {en},
	number = {1},
	urldate = {2026-03-31},
	journal = {Microsystems \& Nanoengineering},
	publisher = {Nature Publishing Group},
	author = {Zou, Haochen and Zhou, Zhibo and Huang, Mengyao and Li, Wenhao and Yang, Bowen and Zhao, Xiao and Li, Ting and Xu, Lijie and Wang, Ting and Wang, Lianhui},
	month = oct,
	year = {2025},
	pages = {191},
}

@article{han_battery-free_2018,
	title = {Battery-free, wireless sensors for full-body pressure and temperature mapping},
	volume = {10},
	url = {https://www.science.org/doi/10.1126/scitranslmed.aan4950},
	doi = {10.1126/scitranslmed.aan4950},
	number = {435},
	urldate = {2026-03-31},
	journal = {Science Translational Medicine},
	publisher = {American Association for the Advancement of Science},
	author = {Han, Seungyong and Kim, Jeonghyun and Won, Sang Min and Ma, Yinji and Kang, Daeshik and Xie, Zhaoqian and Lee, Kyu-Tae and Chung, Ha Uk and Banks, Anthony and Min, Seunghwan and Heo, Seung Yun and Davies, Charles R. and Lee, Jung Woo and Lee, Chi-Hwan and Kim, Bong Hoon and Li, Kan and Zhou, Yadong and Wei, Chen and Feng, Xue and Huang, Yonggang and Rogers, John A.},
	month = apr,
	year = {2018},
	pages = {eaan4950},
}

@article{han_lin_yeap_compact_2025,
	title = {Compact {Environment} in {ISO} 15693 {HF} {RFID} {Tag} {Collision} and {Interference} {With} {KMC}-{PQ}-{RDCR} {Algorithm} {Recovery}},
	volume = {9},
	issn = {2469-7281},
	url = {https://ieeexplore.ieee.org/document/11164331/},
	doi = {10.1109/JRFID.2025.3609858},
	urldate = {2026-03-31},
	journal = {IEEE Journal of Radio Frequency Identification},
	author = {Han Lin Yeap, Herrick and Ying Lim, Le and Seng Eu, Kok and Yap, Kian Meng},
	year = {2025},
	pages = {779--796},
}

@techreport{noauthor_cards_nodate,
	title = {Cards and security devices for personal identification — {Contactless} vicinity objects — {Part} 3: {Anticollision} and transmission protocol},
	shorttitle = {{ISO}/{IEC} 15693-3:2019},
	url = {https://www.iso.org/standard/73602.html},
	language = {en},
	urldate = {2026-03-31},
}

@inproceedings{xie_enabling_2024,
	address = {New York, NY, USA},
	series = {{ACM} {MobiCom} '24},
	title = {Enabling {High}-rate {Backscatter} {Sensing} at {Scale}},
	isbn = {979-8-4007-0489-5},
	url = {https://dl.acm.org/doi/10.1145/3636534.3649351},
	doi = {10.1145/3636534.3649351},
	urldate = {2026-03-30},
	booktitle = {Proceedings of the 30th {Annual} {International} {Conference} on {Mobile} {Computing} and {Networking}},
	publisher = {Association for Computing Machinery},
	author = {Xie, Mingqi and Jin, Meng and Zhu, Fengyuan and Zhang, Yuzhe and Tian, Xiaohua and Wang, Xinbing and Zhou, Chenghu},
	year = {2024},
	pages = {124--138},
}

@inproceedings{takahashi_meander_2022,
	address = {New York, NY, USA},
	series = {{CHI} '22},
	title = {Meander {Coil}++: {A} {Body}-scale {Wireless} {Power} {Transmission} {Using} {Safe}-to-body and {Energy}-efficient {Transmitter} {Coil}},
	isbn = {978-1-4503-9157-3},
	shorttitle = {Meander {Coil}++},
	url = {https://dl.acm.org/doi/10.1145/3491102.3502119},
	doi = {10.1145/3491102.3502119},
	urldate = {2026-03-21},
	booktitle = {Proceedings of the 2022 {CHI} {Conference} on {Human} {Factors} in {Computing} {Systems}},
	publisher = {Association for Computing Machinery},
	author = {Takahashi, Ryo and Yukita, Wakako and Yokota, Tomoyuki and Someya, Takao and Kawahara, Yoshihiro},
	year = {2022},
	pages = {1--12},
}

@inproceedings{sangar_wichronos_2020,
	address = {New York, NY, USA},
	series = {{MobiCom} '20},
	title = {{WiChronos}: energy-efficient modulation for long-range, large-scale wireless networks},
	isbn = {978-1-4503-7085-1},
	shorttitle = {{WiChronos}},
	url = {https://dl.acm.org/doi/10.1145/3372224.3380898},
	doi = {10.1145/3372224.3380898},
	urldate = {2026-03-21},
	booktitle = {Proceedings of the 26th {Annual} {International} {Conference} on {Mobile} {Computing} and {Networking}},
	publisher = {Association for Computing Machinery},
	author = {Sangar, Yaman and Krishnaswamy, Bhuvana},
	year = {2020},
	pages = {1--14},
}

\end{document}